\documentclass[letterpaper]{article}
\usepackage{amssymb,amsmath,amsthm,bbm}

\newcommand{\eq}[1]
    {\begin{equation}
        #1
     \end{equation}}

\newcommand{\spliteq}[1]
    {\begin{equation}\begin{split}
	    #1
	 \end{split}\end{equation}}
	 
\newcommand{\brac}[1]{\left(#1\right)}
\newcommand{\set}[1]{\left\{#1\right\}}
\newcommand{\ebrac}[1]{\left[#1\right]}
\newcommand{\abs}[1]{\left\arrowvert #1\right\arrowvert}

\newcommand{\ket}[1]{\left|#1\right>}
\newcommand{\norm}[1]{\left<#1\left|#1\right>\right.}
\newcommand{\cor}[1]{\left<#1\right>}

\begin{document}

\begin{titlepage}
\setcounter{page}{0}
\begin{flushright}
      hep-th/0702158\\
      ITP--UH-01/07\\
\end{flushright}
\vspace{20mm}
\begin{center}
{\LARGE\bf  On the Various Types of D-Branes in the Boundary $\bf H_3^+$ Model}\\
\vspace{5mm}
{\sc Hendrik Adorf} and {\sc Michael Flohr}\\
\vspace{5mm}
{\sl Institut f\"ur Theoretische Physik,}\\
{\sl Gottfried Wilhelm Leibniz Universit\"at Hannover,}\\
{\sl Appelstra\ss e 2, 30167 Hannover, Germany.}\\
\vspace{5mm}
{\sc e-mail:} \tt adorf, flohr@itp.uni-hannover.de\\
\vspace{5mm}
{\small February 20, 2007}
\end{center}
\vspace{10mm}
\begin{abstract}
\noindent We comment on the D-brane solutions for the boundary $H_3^+$ model that have been proposed so far and point out that many more types of D-branes should be considered possible. We start a systematic derivation of the $1/2$- and $b^{-2}/2$-shift equations corresponding to each type. These equations serve as consistency conditions and we discuss their possible solutions. On this basis, we show for the known $AdS_2^{(d)}$ branes, that only strings transforming in finite dimensional $SL(2)$ representations can couple to them. Moreover, we also demonstrate that strings in the infinite dimensional continuous $SL(2)$ representations do not couple consistently to the known $AdS_2$ branes. For some other types, we show that no consistent solutions seem to exist at all.
\end{abstract}
\vfill
\end{titlepage}

\phantom{A}
\vspace{10mm}
\hrule
\setcounter{section}{-1}
\setcounter{page}{0}
\section{Preamble}
This article is an old version of \cite{Adorf:FactorCnstr} and the reader is requested to read \cite{Adorf:FactorCnstr} instead of this work. Let us briefly explain: In the course of revising the article at hand, it occured to us that some of its statements are misleading, due to some subtleties in the analytic continuations that we have to use. Now, after a thorough revision, our viewpoint on the whole subject has changed and the material has grown immensely. We have therefore decided to publish it as a completely new article \cite{Adorf:FactorCnstr} (which it actually is, since only some parts of the introductory material and of the appendices have stayed unaltered) rather than merely replacing this old one. {\bf This work is therefore superseded by \cite{Adorf:FactorCnstr}}, since \cite{Adorf:FactorCnstr} corrects the misleading formulae and statements and puts everything into a new perspective. Nonetheless, it is our decision to leave the article at hand on the arxive, because the basic ideas of exploring certain patterns systematically and trying to treat the $\rm H_3^+$ boundary two point function analytically are already formulated here.
\vskip .5cm
\hrule

\newpage
\section{Introduction}
The $H_3^+$ model, which is a suggestive way to denote the $SL(2,\mathbb{C})/SU(2)$ WZNW model, has been studied for quite some time now, the motivations being at least fourfold: On the one hand, it falls into the class of noncompact conformal field theories whose general structure and features are very poorly understood so far. On the other hand, it is essential for a study of the bosonic string in certain curved backgrounds. While the $H_3^+$ model itself describes the bosonic string in an euclidean $AdS_3$ background, it can be analytically continued to the lorentzian $AdS_3$ string. The latter is of great interest, particularly in view of the $AdS_3/CFT_2$ correspondence. See references \cite{GKS2}, \cite{KS}, \cite{BORT}, \cite{SAT1}, \cite{SAT2}, \cite{GN}, \cite{GNA}, \cite{SR2}, \cite{MaldaOoguri1}, \cite{MaldaOoguri2}, \cite{MaldaOoguri3} and further references therein. Thirdly, from the euclidean $AdS_3$ string there is a connection to the so-called cigar CFT \cite{RibScho}, which describes a bosonic string moving in an euclidean 2D black hole \cite{WIT}, \cite{DVV}, \cite{BB}, \cite{HPT}. Finally, a forth reason to study the $H_3^+$ model is its very interesting duality to Liouville theory \cite{JT3}, \cite{ZAF}, which has been remarkably generalised only quite recently in \cite{TeschRi}.\\
Concerning the bulk $H_3^+$ model, its structure is apparently quite well explored (see \cite{JT3}, \cite{JT1}, \cite{JT2} and \cite{BP1}), although some subtleties still persist (e.g. \cite{BP1}, \cite{GIRI1}, \cite{GIRI2}, \cite{NIC1}). Looking at the corresponding boundary CFT, we find that the picture is rather more incomplete. In particular, the question of what D-branes can consistently be described does not seem to be fully answered up to now. One approach to this issue, that has been pursued in \cite{PST}, \cite{SR1}, \cite{GKS}, \cite{PS} and \cite{LOP} is to compute boundary one point functions. These are actually fixed to great extent by boundary Ward identities. Their only remaining degree of freedom is the so-called one point amplitude. This is an interesting object to study, because it describes the coupling of a closed string in the bulk to a D-brane. Accordingly, it must depend on the properties of these two objects. Seeing that closed strings are characterised by an $SL(2)$-'spin' label $j$ (see chapter \ref{Review}) and D-branes are labelled by a complex parameter $\alpha$, a one point amplitude is denoted $A(j\arrowvert\alpha)$.\footnote{It can also depend on some other data, see chapters \ref{BraneTypes} and \ref{irrAdS2d-rho2}.} In the sequel, when talking about a D-brane solution, we actually mean a solution for the one point amplitude. For other aspects and further references concerning the boundary $H_3^+$ model, we refer the reader to the lecture notes \cite{VS1} and \cite{VS2}.\\
The strategy in the computation of one point amplitudes is to derive a consistency condition (a so-called shift equation) for them and then try to solve it. The nature of this shift equation is to relate the one point amplitude for some string label $j$ to a sum of one point amplitudes taken at shifted string labels like e.g. $j\pm 1/2$. See equation (\ref{irrAdS2d-rho2-Shift1}) for an example. Generically, solutions do not exist for an arbitrary D-brane label $\alpha\in\mathbb{C}$, but restrictions will apply. By the same token, the labels $j$ of strings that do couple consistently are expected to be constrained.\\
The solution to only one such shift equation is however not unique. Looking at the available results, it is obvious that systematisation and completion are still very much needed (see Table \ref{T1} in chapter \ref{BraneTypes} for an overview of the present situation). What we need to do is
\newcounter{todo}
\begin{list}{\arabic{todo}.)}{\usecounter{todo}\setlength{\topsep 2mm}{\itemsep 0mm}}
\item identify the mechanisms that give rise to different types of D-brane solutions and then explore their consequences in a systematic fashion.\par
\item derive more shift equations in order to extract unique solutions. It is very likely that this will call for modifications of the solutions that have been proposed so far.
\end{list}
Let us explain the first point in more detail. In \cite{PST}, the authors showed that there are two classes of D-branes: $AdS_2$ and $S^2$ branes.\footnote{This seems to contradict the fact that there are actually {\sl four} different gluing conditions (see chapter \ref{GlueConds}), resulting in four different classes of D-branes. However, some of these four might be isomorphic, giving rise to identical one point amplitudes. We will actually see this explicitely for two different gluing conditions in chapters \ref{irrAdS2d-rho2} and \ref{irrAdS2d-rho1} and in chapter \ref{irrAdS2c-rho2-1}.} They derived one shift equation for each class and also proposed solutions. Afterwards, \cite{SR1} enlarged the picture and introduced the so-called $AdS^{(d)}_2$ branes, $(d)$ standing for {\sl discrete}. The author of \cite{SR1} was guided by some relation between the ZZ and FZZT branes of Liouville theory that, in the spirit of the Liouville/$H_3^+$ correspondence of \cite{TeschRi}, was carried over to the $AdS_2$ branes of \cite{PST}. However, these new D-branes can also be understood as arising from a substantial difference in the derivation of the shift equation. In that derivation, a special two point function involving one degenerate field is considered (see chapter \ref{Review} for an explanation of the term 'degenerate field'). The benefit of using a degenerate field here is, that it allows to solve for the two point function exactly. In order to get a shift equation for the one point amplitude, two different treatments are then possible: The degenerate field can either be taken to approach the boundary and be expanded in terms of boundary fields, or the two fields can be taken far apart from each other, such that the two point function factorises into a product of two one point functions. The first case results in the $AdS_2$, whereas the second case leads to the $AdS_2^{(d)}$ shift equations. This treatment can always be applied, no matter what gluing condition we are using. This has actually been recognised, but not fully exploited, by the authors of \cite{GKS}.\\
On top of that, there might be even more D-brane solutions. In \cite{GKS}, a solution to the boundary conformal Ward identities for the one point function, that is everywhere regular in the internal variable (see chapter \ref{Review}), was proposed. Opposed to this solution, \cite{PST}, \cite{SR1} and \cite{LOP} use a one point function that is not everywhere regular. While both solutions are correct (see chapter \ref{BraneTypes}), we find that they give rise to slightly different shift equations (see chapters \ref{irrAdS2d-rho1}, \ref{regAdS2d-rho2} and \ref{regAdS2d-rho1} in case of the discrete and \ref{irrAdS2c-rho2-1}, \ref{regAdS2c-rho2} and \ref{regAdS2c-rho1} for the continuous D-branes). The modifications that arise for the regular dependence opposed to the irregular one, change the qualitative behaviour of possible solutions significantly. Consequently, not only should one distinguish between continuous and discrete, but also between regular\footnotemark and irregular\addtocounter{footnote}{-1}\footnote{We are going to introduce this terminology in chapter \ref{BraneTypes}.} D-brane solutions. In order to be sure that no subtleties have been overlooked, the two different types of shift equation, discrete and continuous, should be derived systematically for each gluing condition, with regular as well as irregular dependence.\\
The plan of this paper is as follows: After having fixed some notation in chapter \ref{Review}, we elaborate on the problems mentioned above in chapters \ref{GlueConds} and \ref{BraneTypes}. This is followed by the derivation of the shift equation involving degenerate field $\Theta_{b^{-2}/2}$ for an irregular $AdS_2^{(d)}$ brane in chapter \ref{irrAdS2d-rho2}. There, we also show that the solution that had been proposed earlier does only solve this new shift equation under the very special condition that the field transforms in a finite dimensional $SL(2)$ representation. Then we go on to derive and discuss the solutions to the shift equations (involving degenerate fields with $SL(2)$-'spin' labels $j=1/2$ and $j=b^{-2}/2$ respectively) for the remaining discrete $AdS_2$ branes in chapters \ref{irrAdS2d-rho1}, \ref{regAdS2d-rho2} and \ref{regAdS2d-rho1}. Afterwards, in chapters \ref{irrAdS2c-rho2-1}, \ref{regAdS2c-rho2} and \ref{regAdS2c-rho1}, we give the $1/2$- and $b^{-2}/2$-shift equations for the various continuous $AdS_2$ branes and also comment on their possible solutions. Here, we can show that fields with labels $j\in -1/2+i\mathbb{R}$ can not couple consistently to the irregular $AdS_2^{(c)}$ branes of \cite{PST} and \cite{LOP}. Finally, we summarise our results in chapter \ref{Conclusion}, where we also suggest further directions and discuss open questions. The more technical calculational details and some useful formulae are found in the appendices.

\section{\label{Review}A Brief Review of the Bulk $\bf H_3^+$ Model}
The bulk $H_3^+$ model has been fairly well studied, see \cite{JT3}, \cite{JT1}, \cite{JT2} and \cite{BP1}. Here, we essentially fix our notation (which follows very closely \cite{PST}) and summarise those facts and formulae which will be indispensable in the sequel. They can all be found in \cite{JT1}, \cite{JT2} and \cite{PST}.\\
Besides conformal symmetry, the $H_3^+$ model possesses an affine $\hat{sl}(2,\mathbb{C})_k\times\hat{sl}(2,\mathbb{C})_k$ symmetry, i.e. its chiral algebra does not only consist of an energy momentum tensor $T(z)$, but also of the currents $J^a(z)=\sum_n z^{-n-1}J^a_n$, $a\in\set{+,-,3}$ (plus a corresponding antichiral sector). Primary fields fall into representations of the zero mode algebra (generated by the operators $J^a_0$) and are henceforth labelled by a pair of $sl(2,\mathbb{C})$-'spins' $(j,\bar{j})$, and a pair of internal variables, which will be denoted by $(u,\bar{u})\in\mathbb{C}^2$, so that a typical primary field should be denoted $\Theta_{j,\bar{j}}(u,\bar{u}|z,\bar{z})$. However, from now on we will always suppress the barred variables. The $\hat{sl}(2,\mathbb{C})_k$-currents act on these primaries via the operator product expansion (OPE)
\eq{J^a(z)\Theta_j(u|w)=\frac{D^a_j(u)\phi_j(u|w)}{z-w},}
i.e. the zero mode algebra is represented through the differential operators $D^a_j(u)$, given by
\eq{
D^+_j(u):=-u^2\partial_u+2ju, \hskip .5cm 
D^-_j(u):=\partial_u, \hskip .5cm 
D^3_j(u):=u\partial_u-j.}
Analogous formulae hold for the antichiral sector. Through the standard Sugawara construction, the energy momentum tensor is expressed in terms of products of the currents and thereby a relation between conformal weight $h$ and 'spin'-label $j$ of primary fields is established:
\eq{h\equiv h(j)=-\frac{j(j+1)}{k-2}=:-b^2 j(j+1).}
Note that there is a reflection symmetry, namely $h(-j-1)=h(j)$. This leads one to identify the representations with labels $j$ and $-j-1$ and gives rise to a relation between primary fields $\Theta_j(u|z)$ and $\Theta_{-j-1}(u|z)$:
\eq{\label{RefSymm}\Theta_j(u|z)=-R(-j-1)\frac{2j+1}{\pi}\int_{\mathbb{C}}{\rm d}^2 u'|u-u'|^{4j}\Theta_{-j-1}(u'|z),}
where the {\sl reflection amplitude} $R(j)$ is given by
\eq{\label{Rj}R(j)=-\nu_b^{2j+1}\frac{\Gamma(1+b^2(2j+1))}{\Gamma(1-b^2(2j+1))}.}
The physical spectrum (normalisable operators) of the bulk theory consists of the so-called continuous representations \cite{JT2}, that are parametrised through $j\in -\frac{1}{2}+i\mathbb{R}_+$ and are in fact infinite dimensional.\\
By the usual operator-state correspondence, to each primary field $\Theta_j$ corresponds a highest weight state $\ket{j}$. It has the property that $J^a_n\ket{j}=0$ for all $n>0$. Acting on it with the $J^a_{n<0}$ generates a whole Verma module $V_j$. These modules are reducible, iff
\eq{j=j_{r,s}:=-\frac{1}{2}+\frac{1}{2}r+\frac{b^{-2}}{2}s,}
where either $r\geq 1$, $s\geq 0$ or $r<-1$, $s<0$ (see \cite{JT1}). This means that they possess null-submodules. These are submodules that are generated by so-called {\sl null states} (or {\sl singular vectors}), i.e. states $\ket{\mathrm{null}}$ with $\norm{\mathrm{null}}=0$. Those primary fields $\Theta_{j_{r,s}}$ that give rise to reducible modules are called {\sl degenerate fields}. In order to get an irreducible module out of a reducible one, all null-submodules have to be divided out of the original module. This in turn gives rise to certain differential equations, that all correlators involving the corresponding degenerate field have to solve.

\section{Boundary $\bf H_3^+$}
In this section, we have several comments to make on the boundary CFT that one obtains from the $H_3^+$ model. Then, we give the details of our derivation of the $b^{-2}/2$ shift equation for the irregular discrete $AdS_2$ branes, that have been studied in \cite{SR1}. With the help of our new shift equation, we show that the solution proposed in \cite{SR1} is only valid for strings that transform in finite dimensional $SL(2)$ representations. 

\subsection{\label{GlueConds}Gluing Conditions}
We choose maximal symmetry preserving boundary conditions. This is done by imposing a gluing condition along the boundary (which is taken to be the real axis)
\eq{J^a(z)-\rho(\bar{J}^a(\bar{z}))=0 \hskip .5cm {\rm at} \hskip .1cm z=\bar{z},}
where $\rho$ is the 'gluing map' i.e. an automorphism of the chiral algebra which leaves the Virasoro field invariant. Thus, by the Sugawara construction, we also have
\eq{T(z)+\bar{T}(\bar{z})=0 \hskip .5cm {\rm at} \hskip .1cm z=\bar{z},}
and hence not only is a subgroup of the current algebra symmetry preserved, but also half of the conformal symmetry. In the case of $SL(2)$ there are four possible gluing maps $\rho_1,\dots,\rho_4$\footnote{At first sight, one might be tempted to see four more automorphisms, like e.g. $\rho(\bar{J}^{3})=\bar{J}^{3}, \rho(\bar{J}^{\pm})=\bar{J}^{\mp}$, and so on, but it turns out that with these, the modes of the symmetry currents that leave the boundary state invariant do not close to form an algebra. Furthermore, the Ward identites that were associated to such boundary conditions do not have any solutions.}: 
\spliteq{
\rho_1(\bar{J}^3)=\bar{J}^3 \hskip .5cm 
&\rho_1(\bar{J}^{\pm})=\bar{J}^{\pm},\\
\rho_2(\bar{J}^3)=\bar{J}^3 \hskip .5cm 
&\rho_2(\bar{J}^{\pm})=-\bar{J}^{\pm},\\
\rho_3(\bar{J}^3)=-\bar{J}^3 \hskip .5cm
&\rho_3(\bar{J}^{\pm})=\bar{J}^{\mp},\\
\rho_4(\bar{J}^3)=-\bar{J}^3 \hskip .5cm
&\rho_4(\bar{J}^{\pm})=-\bar{J}^{\mp}.\\
}                                                         
For now, we will only be concerned with the first and second case, $\rho_1$ and $\rho_2$, and the associated branes are conventionally called $AdS_2$ D-branes \cite{PST}.

\subsection{\label{BraneTypes}Various Types of D-Branes}
For each of the above four classes of boundary conditions, one can obtain at least two different D-brane solutions: The 'continuous' and the 'discrete' D-branes. By the term 'D-brane solution' we mean the one point amplitude of a generic field $\Theta_j$ in the presence of some boundary condition. The characterising adjectives 'continuous' and 'discrete' relate to the parameter spaces of these solutions. For example, in \cite{PST}, a solution for the {\sl continuous} $AdS_2$ branes was proposed, whereas \cite{SR1} proposed a solution for the {\sl discrete} $AdS_2$ branes. From now on, we will carefully distinguish these different kinds of solutions, by adding a superscript $(c)$ in case of a continuous brane and $(d)$ for a discrete one, as it has already been done in \cite{SR1}.\\ 
Moreover, we want to argue that there are even more possible D-brane solutions, that are distinguished by their regularity behaviour when approaching the boundary in internal $u$-space. Let us explain in detail why this is the case for the example that the gluing map is $\rho=\rho_2$ (the other cases can clearly be treated in just the same way). It is the Ward identites that fix the $u$-dependence of the one point function $G^{(1)}_{j,\alpha}(u|z):=\cor{\Theta_j(u|z)}_{\alpha}$ in the presence of boundary condition $\alpha$ entirely. The equation for $J^-$ tells us that it is a function of $u+\bar{u}$ only. The equations for $J^3$ and $J^+$ have a singularity at $0=u+\bar{u}=:2u_1$, hence we have to distinguish two cases. The solution for $u_1>0$ is 
\eq{G^{(1)}_{j,\alpha}(u;u_1>0|z)=(u+\bar{u})^{2j}A^{+}_{j,\alpha}(z)} 
and the one for $u_1<0$ reads 
\eq{G^{(1)}_{j,\alpha}(u;u_1<0|z)=(u+\bar{u})^{2j}A^{-}_{j,\alpha}(z).} 
But notice that we could have equally well written 
\eq{G^{(1)}_{j,\alpha}(u;u_1<0|z)=|u+\bar{u}|^{2j}\tilde{A}^{-}_{j,\alpha}(z),} 
where we have just redefined the "constant": $\tilde{A}^{-}_{j,\alpha}(z)=(-)^{2j}A^{-}_{j,\alpha}(z)$.
\begin{table}
\begin{tabular}{|c||c|c|c|c|c|} \hline
             &$u$-dependence   
                       &\multicolumn{2}{c|}{shift equation (continuous)}   
                                 &\multicolumn{2}{c|}{shift equation (discrete)}\\
             &         
                       &for $\Theta_{1/2}?$    &for $\Theta_{b^{-2}/2}?$ 
                                 &for $\Theta_{1/2}?$     &for $\Theta_{b^{-2}/2}?$\\ \hline\hline
$\rho_1$     &$|u-\bar{u}|^{2j}$ 
                       &\cite{LOP}             &---
                                 &---                    &---\\ \cline{2-6}
             &$(u-\bar{u})^{2j}$ 
                       &\cite{GKS}             &---
                                 &\cite{GKS}             &\cite{GKS}\\ \hline
$\rho_2$     &$|u+\bar{u}|^{2j}$ 
                       &\cite{PST}             &---                 
                                 &\cite{SR1}             &---\\ \cline{2-6}
             &$(u+\bar{u})^{2j}$ 
                       &---                    &---
                                 &---                    &---\\ \hline
$\rho_3$     &$|-1+u\bar{u}|^{2j}$ 
                       &---                    &---                   
                                 &---                    &---\\ \cline{2-6}  
             &$(-1+u\bar{u})^{2j}$       
                       &---                    &---
                                 &\cite{GKS}             &\cite{GKS}\\ \hline
$\rho_4$     &$(1+u\bar{u})^{2j}$       
                       &---                    &---
                                 &\cite{PST}             &---\\ \hline                      
\end{tabular}
\caption{\label{T1} Classes of D-brane solutions and status of their exploration. \cite{GKS} did not distinguish between amplitudes $A^-$ and $A^+$, which is however inevitable (see text). We are therefore reconsidering their results. Note that only one version of $u$-dependence appears for $\rho_4$, as the expression is always strictly positive.}
\end{table}
This seems like a harmless thing to do, but we need to be aware that the $u$ dependence has changed from being regular at $u_1=0$ to irregular. In this and the next chapter, we will compute the one point amplitudes resulting from both these ans\"atze and find that they are indeed very different in nature. The corresponding D-branes will be called {\sl regular} or {\sl irregular}, respectively. Whether this is an appropriate and useful nomination remains to be seen.\\
It should be mentioned that in the literature, both kinds of solutions, regular and irregular ones, have been studied. For example, \cite{PST} and \cite{LOP} look at irregular $AdS^{(c)}_2$ and \cite{SR1} treats irregular $AdS^{(d)}_2$ branes, whereas \cite{GKS} studies regular solutions. But up to now, at least to our knowledge, nobody has pointed out that for {\sl every} case of boundary condition $\rho_1,\dots\rho_4$, we should actually look for {\sl both} kinds of solutions. Table \ref{T1} shows how little of the 'landscape' has actually been explored so far.\\
Summarising, we want to say that a variety of different D-brane solutions has been proposed, but they are scattered, written in varying notation and conventions, and we are far from a systematic analysis that distinguishes carefully different classes of solutions and tabulates them in a unified notation. See again Table \ref{T1} for a quick overview of the present situation. Furthermore, except for one case in \cite{GKS}, it has always been only {\sl one} consistency condition on which the proposed solutions were based, namely the shift equation for the degenerate field $\Theta_{1/2}$. The solutions to this equation are not unique and at least a second consistency condition should be derived, that can fix the solution uniquely. The shift equation for the degenerate field $\Theta_{b^{-2}/2}$ does this job. It is our goal to achieve completion of the aforementioned classifying table of solutions (Table \ref{T1}) and fix the solutions uniquely through usage of both shift equations. This paper is an incomplete step into that direction, but we are confident to complete it in the near future.\\
Finally, it seems needless to say that even our classifying table will not establish the full consistency of solutions, because in principle there are infinitely many shift equations, arising from infinitely many degenerate fields. Nevertheless we find it useful to tabulate the solutions to the two simplest constraints as a first step, because it clarifies what we have to look at and can possibly already exclude some cases.

\subsection{\label{irrAdS2d-rho2}Irregular $AdS^{(d)}_2$ Branes - Gluing Map $\rho_2$}
\subsubsection{Shift Equations for the Boundary One Point Amplitudes}
The gluing map is $\rho_2$. Choosing the irregular $u$-dependence, it restricts the one point function in the presence of boundary condition $\alpha$ to be of the form
\eq{\cor{\Theta_j(u\arrowvert z)}_{\alpha}=\abs{z-\bar{z}}^{-2h(j)}\abs{u+\bar{u}}^{2j}A_{\sigma}(j\arrowvert\alpha).}
The unknown function $A_{\sigma}(j\arrowvert\alpha)$ is the {\sl one point amplitude}. Note that it still depends on $\sigma:=\mathrm{sgn}(u+\bar{u})$. Its physical interpretation is that it describes the strength of coupling of a closed string with label $j$ to the brane labelled by $\alpha$. It is possible to obtain necessary conditions on $A_{\sigma}(j\arrowvert \alpha)$ by considering two point functions involving a degenerate field. This strategy has been pursued in \cite{PST}, \cite{GKS} and \cite{LOP} for degenerate field $\Theta_{1/2}$ (and in \cite{GKS} one case has also been treated using the degenerate field $\Theta_{b^{-2}/2}$, see Table \ref{T1}). However, only a few cases have been checked so far and refering once again to Table \ref{T1}, it becomes clear that lots of constraints (shift equations) remain to be computed. Many shift equations that we derive in this paper will be given for the first time.\\
Let us now illustrate the whole procedure for the irregular $AdS^{(d)}_2$ branes in case of a two point function involving the degenerate field $\Theta_{b^{-2}/2}$. This will lead us to a formerly unknown shift equation.\\
Using the Ward identities, the form of the two point function $G^{(2)}_{j,\alpha}(u_i\arrowvert z_i):=\cor{\Theta_{b^{-2}/2}(u_2\arrowvert z_2)\Theta_j(u_1\arrowvert z_1)}_{\alpha}$ can be partially fixed as
\spliteq{\label{B2pt}G^{(2)}_{j,\alpha}(u_1,u_2\arrowvert z_1,z_2)=&\abs{z_1-\bar{z}_1}^{2[h(b^{-2}/2)-h(j)]}\abs{z_1-\bar{z}_2}^{-4h(b^{-2}/2)}\cdot\\
&\cdot \abs{u_1+\bar{u}_1}^{2j-b^{-2}}\abs{u_1+\bar{u}_2}^{2b^{-2}}H^{(2)}_{j,\alpha}(u\arrowvert z),} 
where $H^{(2)}_{j,\alpha}(u\arrowvert z)$ is an unknown function of the crossing ratios 
\eq{z:=\frac{\abs{z_2-z_1}^2}{\abs{z_2-\bar{z}_1}^2} \hskip .5cm {\rm and} \hskip .5cm
u:=\frac{\abs{u_2-u_1}^2}{\abs{u_2+\bar{u}_1}^2}.}
Now, the standard Knizhnik-Zamolodchikov equations are used to deduce a partial differential equation for $H^{(2)}_{j,\alpha}(u\arrowvert z)$ (see Appendix \ref{KniZa}). Since one field operator is the degenerate field $\Theta_{b^{-2}/2}$, its space of solutions is finite dimensional, in fact it consists of three conformal blocks only, namely those for $j_{\pm}:=j\pm b^{-2}/2$ and $j_x:=-j-1-b^{-2}/2$. Hence, the general solution reads
\eq{\label{H2}H^{(2)}_{j,\alpha}(u\arrowvert z)=\sum_{\epsilon=+,-,x} a^{j}_{\epsilon}(\alpha){\cal F}^{s}_{j,\epsilon}(u\arrowvert z),}
where the conformal blocks ${\cal F}^{s}_{j,\epsilon}(u\arrowvert z)$ are given in Appendix \ref{KniZa}, and the $a^{j}_{\epsilon}(\alpha)$ are some still undetermined coefficients. They are fixed by using the bulk OPE of the two field operators on the L.H.S. and taking the appropriate limit $\abs{z_2-z_1}\rightarrow 0$ on the R.H.S. of (\ref{B2pt}). The $a^{j}_{\epsilon}(\alpha)$ will then generically turn out to be some product of bulk OPE coefficient times one point amplitude, which is why the $\alpha$-dependence occurs in the $a^{j}_{\epsilon}$-coefficients. We find (see Appendix \ref{LinComb} for details) that 
\eq{\label{a-coeffs}a^{j}_{\epsilon}(\alpha)=C_{\epsilon}(j)A_{\sigma}(j_{\epsilon}\arrowvert\alpha).}
where the $C_{\epsilon}(j)$ are bulk OPE coefficients, see Appendix \ref{OPE2} for their explicit expressions. The boundary two point function (\ref{B2pt}) is now determined exactly.\\
In order to get a shift equation for the {\sl discrete} brane solution, we take the limit $\abs{z_2-z_1}\rightarrow\infty$ ($\Rightarrow z\rightarrow 1$) followed by $\abs{u_2-u_1}\rightarrow\infty$ ($\Rightarrow u\rightarrow 1$). Upon doing this, the L.H.S. of (\ref{B2pt}) turns into a product of two one point functions, by cluster decomposition. Hence, we get
\spliteq{G^{(2)}_{j,\alpha}(u_1,u_2\arrowvert z_1,z_2)\simeq & \abs{z_1-\bar{z_1}}^{-2h(j)}\abs{z_2-\bar{z_2}}^{-2h(b^{-2}/2)}\cdot\\
& \cdot\abs{u_1+\bar{u_1}}^{2j}\abs{u_2+\bar{u_2}}^{b^{-2}}A_{\sigma}(j\arrowvert\alpha)A_{\sigma}(b^{-2}/2\arrowvert\alpha).}
On the R.H.S. of (\ref{B2pt}), we just take our exact expression (involving the results (\ref{H2}) and (\ref{a-coeffs})) and perform the limit explicitly. For details, see Appendix \ref{InftyLim}. If we redefine the one point amplitude (see Appendix \ref{RefCnstr} for a motivation of {\sl this} particular redefinition)
\eq{\label{ReDef}f_{\sigma}(j)\equiv f_{\sigma}(j\arrowvert\alpha):=\nu_b^j\Gamma(1+b^2(2j+1))A_{\sigma}(j\arrowvert\alpha)} 
and equate the two expressions from L.H.S. and R.H.S., we arrive at our new additional shift equation for the irregular $AdS^{(d)}_2$ brane:
\spliteq{\label{irrAdS2d-rho2-Shift2}\frac{e^{-i\pi\brac{j+\frac{b^{-2}}{2}}}}{\Gamma(1+b^2)}f_{\sigma}\brac{\frac{b^{-2}}{2}}f_{\sigma}(j)&=e^{-i\pi\brac{j+\frac{b^{-2}}{2}}}f_{\sigma}\brac{j+\frac{b^{-2}}{2}}-\\
-e^{i\pi\brac{j+\frac{b^{-2}}{2}}}f_{-\sigma}&\brac{j+\frac{b^{-2}}{2}}+e^{-i\pi\brac{j-\frac{b^{-2}}{2}}}f_{\sigma}\brac{j-\frac{b^{-2}}{2}}.}
For completeness let us also write down the formerly known shift equation \cite{SR1} for the redefined one point amplitude (\ref{ReDef}). It is 
\spliteq{\label{irrAdS2d-rho2-Shift1}-\frac{1}{\pi}\Gamma(-b^2)\sin[2\pi b^2]\sin[\pi b^2(2j+1)]f_{\sigma}\brac{\frac{1}{2}}f_{\sigma}\brac{j}=&\\
\sin[\pi b^2(2j+2)]f_{\sigma}\brac{j+\frac{1}{2}}-\sin[&\pi b^2 2j]f_{\sigma}\brac{j-\frac{1}{2}}.}

\subsubsection{Solving the Shift Equations}
The formlery know shift equation (\ref{irrAdS2d-rho2-Shift1}) is solved by \cite{SR1}
\eq{\label{irrAdS2d-rho2-Sol}f_{\sigma}(j\arrowvert m,n)=\frac{i\pi\sigma e^{i\pi m}}{\Gamma(-b^2)\sin[\pi nb^2]}e^{-i\pi\sigma(m-\frac{1}{2})(2j+1)}\frac{\sin[\pi nb^2(2j+1)]}{\sin[\pi b^2(2j+1)]},}
with $n,m\in\mathbb{Z}$.\footnote{This is how the solution has been given in \cite{SR1}. In fact we only need $m\in\mathbb{Z}$ here. $n\in\mathbb{Z}$ is required later, in the $b^{-2}/2$-shift equation.} Note that this also satisfies the reflection symmetry constraint (\ref{RefSymmCnstr-irr}). But now, using our second shift equation (\ref{irrAdS2d-rho2-Shift2}), one can check that this is only a solution to {\sl both} shift equations for 
\eq{j\in\frac{1}{2}\mathbb{Z},} 
i.e. only strings in finite (nonunitary) representations would couple to these branes. Of course, the question persists if there is a different solution that is valid for all $j$.

\section{\label{Discrete}More on the Discrete D-Branes}
This section is more or less a collection of new shift equations and their solutions, that we have derived for a variety of different cases. Apart from some tedious but yet important details, especially involving signs, the calculations go as in chapter \ref{irrAdS2d-rho2}, which is why we do not go into great detail here.

\subsection{\label{irrAdS2d-rho1}Irregular $AdS_2^{(d)}$ Branes - Gluing Map $\rho_1$}
\subsubsection{Shift Equations}
Choosing the irregular $u$-dependence, the gluing map $\rho_1$ restricts the one point function in the presence of boundary condition $\alpha$ to be of the form
\eq{\cor{\Theta_j(u\arrowvert z)}_{\alpha}=\abs{z-\bar{z}}^{-2h(j)}\abs{u-\bar{u}}^{2j}A_{\sigma}(j\arrowvert\alpha).}
Our ansatz for the boundary two point function with degenerate field $t/2$, $t=1,b^{-2}$ (fixing the $u_i$ and $z_i$ dependence up to a dependence on the crossing ratios) is
\spliteq{G^{(2)}_{j,t,\alpha}(u_1,u_2\arrowvert z_1,z_2)=\abs{z_1-\bar{z}_1}&^{2[h(t/2)-h(j)]}\abs{z_1-\bar{z}_2}^{-4h(t/2)}\cdot\\
&\cdot \abs{u_1-\bar{u}_1}^{2j-t}\abs{u_1-\bar{u}_2}^{2t}H^{(2)}_{j,t,\alpha}(u\arrowvert z),} 
with crossing ratios
\eq{z:=\frac{\abs{z_2-z_1}^2}{\abs{z_2-\bar{z}_1}^2} \hskip .5cm {\rm and} \hskip .5cm
u:=\frac{\abs{u_2-u_1}^2}{\abs{u_2-\bar{u}_1}^2}.}
The conformal blocks that solve the Knizhnik-Zamolodchikov equations turn out to be just the same ones as for gluing map $\rho_2$, so for $t=b^{-2}$ they are given by (\ref{ConfBlocks}) with parameters
\eq{\alpha=\beta=-b^{-2},\hskip .3cm \beta'=-2j-1-b^{-2},\hskip .3cm \gamma=-2j-b^{-2}}
and for $t=1$ see \cite{PST}.
Also, in both cases ($t=1,b^{-2}$), the expansion coefficients stay as before:
\eq{a^{j}_{\epsilon}(\alpha)=C_{\epsilon}(j)A_{\sigma}(j_{\epsilon}\arrowvert\alpha).}
Taking the limit $\abs{z_2-z_1}\rightarrow\infty$ followed by the same in the $u$'s, we obtain the same shift equations as for gluing map $\rho_2$, namely
\spliteq{\label{irrAdS2d-rho1-Shift1}-\frac{1}{\pi}\Gamma(-b^2)\sin[2\pi b^2]\sin[\pi b^2(2j+1)]f_{\sigma}\brac{\frac{1}{2}}f_{\sigma}\brac{j}=&\\
\sin[\pi b^2(2j+2)]f_{\sigma}\brac{j+\frac{1}{2}}-\sin[&\pi b^2 2j]f_{\sigma}\brac{j-\frac{1}{2}}}
and
\spliteq{\label{irrAdS2d-rho1-Shift2}\frac{e^{-i\pi\brac{j+\frac{b^{-2}}{2}}}}{\Gamma(1+b^2)}f_{\sigma}\brac{\frac{b^{-2}}{2}}f_{\sigma}(j)&=e^{-i\pi\brac{j+\frac{b^{-2}}{2}}}f_{\sigma}\brac{j+\frac{b^{-2}}{2}}-\\
-e^{i\pi\brac{j+\frac{b^{-2}}{2}}}f_{-\sigma}&\brac{j+\frac{b^{-2}}{2}}+e^{-i\pi\brac{j-\frac{b^{-2}}{2}}}f_{\sigma}\brac{j-\frac{b^{-2}}{2}}.}
This means that the irregular discrete D-branes that arise from gluing maps $\rho_1$ and $\rho_2$ respectively, are indeed isomorphic. Compare to our remarks in the introduction, where we explained that this is likely to happen.

\subsubsection{Solving the Shift Equations}
See the corresponding section in chapter \ref{irrAdS2d-rho2}, since the shift equations are identical. For convenience, we state the result here again. A solution to both shift equations and the reflection sysmmetry constraint is given, for $j\in\frac{1}{2}\mathbb{Z}$, by
\eq{\label{irrAdS2d-rho1-Sol}f_{\sigma}(j\arrowvert m,n)=\frac{i\pi\sigma e^{i\pi m}}{\Gamma(-b^2)\sin[\pi nb^2]}e^{-i\pi\sigma(m-\frac{1}{2})(2j+1)}\frac{\sin[\pi nb^2(2j+1)]}{\sin[\pi b^2(2j+1)]},}
with $n,m\in\mathbb{Z}$. Of course, it is again an open problem whether a solution valid for all $j$ exists.

\subsection{\label{regAdS2d-rho2}Regular $AdS_2^{(d)}$ Branes - Gluing Map $\rho_2$}
\subsubsection{Shift Equations}
This time choosing the regular $u$-dependence, the gluing map $\rho_2$ fixes the one point function as
\eq{\cor{\Theta_j(u\arrowvert z)}_{\alpha}=\brac{z-\bar{z}}^{-2h(j)}\brac{u+\bar{u}}^{2j}A_{\sigma}(j\arrowvert\alpha).}
The boundary two point function with degenerate field $t/2$, $t=1,b^{-2}$ is
\spliteq{G^{(2)}_{j,t,\alpha}(u_1,u_2\arrowvert z_1,z_2)=\brac{z_1-\bar{z}_1}&^{-2h(j)}\brac{z_2-\bar{z}_2}^{-2h(t/2)}\cdot\\
&\cdot \brac{u_1+\bar{u}_1}^{2j}\brac{u_2+\bar{u}_2}^{t}H^{(2)}_{j,t,\alpha}(u\arrowvert z),} 
with crossing ratios
\eq{z:=\frac{\abs{z_1-z_2}^2}{\brac{z_1-\bar{z}_1}\brac{z_2-\bar{z}_2}} \hskip .5cm {\rm and} \hskip .5cm
u:=-\frac{\abs{u_1-u_2}^2}{\brac{u_1+\bar{u}_1}\brac{u_2+\bar{u}_2}}.}
Solving the Knizhnik-Zamolodchikov equations for $t=1$ results in the following conformal blocks:
\spliteq{\label{ConfBlocks-1/2}{\cal F}^{s}_{+,j}(u\arrowvert z)&=z^{-b^{2}j}(1-z)^{-b^{2}j}\set{F(a,b;c\arrowvert z)-u\brac{\frac{b}{c}}F(a,b+1;c+1\arrowvert z)},\\
{\cal F}^{s}_{-,j}(u\arrowvert z)&=\left. z^{b^2(j+1)}(1-z)^{b^{2}j}\right\{uF(a-c,b-c+1;1-c\arrowvert z)-\\
&-\left. z\brac{\frac{a-c}{1-c}}F(a-c+1,b-c+1;2-c\arrowvert z)\right\}}
($F(a,b;c\arrowvert z)$ is the Hypergeometric function), with parameters 
\eq{a=-b^2(2j+2),\hskip .3cm b=-b^2(2j),\hskip .3cm c=-b^2(2j+1).}
The solution for $t=b^{-2}$ is again provided by the conformal blocks (\ref{ConfBlocks}), but this time with parameters
\eq{\alpha=-2j,\hskip .3cm \beta=-b^{-2},\hskip .3cm \beta'=-2j-1-b^{-2},\hskip .3cm \gamma=-2j-b^{-2}.}
Also the expansion coefficients have to be modified slightly in this case. For $t=1$, we have
\eq{a^{j,1/2}_{\epsilon}(\alpha)=\epsilon C^{1/2}_{\epsilon}(j)A_{\sigma}(j_{\epsilon}\arrowvert\alpha)}
and for $t=b^{-2}$
\spliteq{a^{j,b^{-2}/2}_{+}(\alpha)&=C^{b^{-2}/2}_{+}(j)A_{\sigma}(j_{+}\arrowvert\alpha),\\
a^{j,b^{-2}/2}_{-}(\alpha)&=e^{i\pi b^{-2}}C^{b^{-2}/2}_{-}(j)A_{\sigma}(j_{-}\arrowvert\alpha),\\
a^{j,b^{-2}/2}_{x}(\alpha)&=-e^{2\pi i(j+b^{-2}/2)}C^{b^{-2}/2}_{x}(j)A_{\sigma}(j_{x}\arrowvert\alpha).}
Taking the limit $\abs{z_2-z_1}\rightarrow\infty$ followed by the same in the $u$'s, we have this time that $u,z\rightarrow -\infty$. Therefore, we have to take different analytic continuations of the occuring Hypergeometric and Appell functions than before. See appendices \ref{Hypergeo} and \ref{Appell}
or consult the books \cite{Exton} and \cite{Bateman}. We obtain the followong shift equations
\spliteq{\label{regAdS2d-rho2-Shift1}-\frac{1}{\pi}\Gamma(-b^2)\sin[2\pi b^2]\sin[\pi b^2(2j+1)]f_{\sigma}\brac{\frac{1}{2}}f_{\sigma}(j)&=\\
e^{-i\pi b^{2}j}\sin[\pi b^2(2j+2)]f_{\sigma}\brac{j+\frac{1}{2}}+e^{i\pi b^{2}j}&\sin[\pi b^2 2j]f_{\sigma}\brac{j-\frac{1}{2}}}
and
\spliteq{\label{regAdS2d-rho2-Shift2}\frac{e^{-i\pi(j+b^{-2})}}{\Gamma(1+b^2)}f_{\sigma}\brac{\frac{b^{-2}}{2}}f&_{\sigma}(j)=e^{-2\pi i\brac{j+\frac{b^{-2}}{2}}}f_{\sigma}\brac{j+\frac{b^{-2}}{2}}-\\
&-e^{4\pi i\brac{j+\frac{b^{-2}}{2}}}f_{-\sigma}\brac{j+\frac{b^{-2}}{2}}-f_{\sigma}\brac{j-\frac{b^{-2}}{2}}.}

\subsubsection{Solving the Shift Equations}
In a first step, we solve the $1/2$-shift equation (\ref{regAdS2d-rho2-Shift1}) together with the reflection symmetry constraint (\ref{RefSymmCnstr-reg}). We find the solution
\eq{\label{regAdS2d-rho2-Sol}f_{\sigma}(j\arrowvert m)=-2\pi\sigma\sin[\pi b^2/2]\frac{e^{-i\pi\frac{3b^2}{4}}}{\Gamma(-b^2)}e^{i\pi m}e^{-i\pi\sigma(m-\frac{1}{2})(2j+1)}\frac{\exp\ebrac{i\pi\frac{b^2}{4}(2j+1)^2}}{\sin[\pi b^2(2j+1)]},}
with $m\in\mathbb{Z}$. This seems to be the most general adaption of (\ref{irrAdS2d-rho2-Sol}) to the case at hand. But note that this time the $\sin[2j+1]$ behaviour is replaced by an $\exp[2j+1]^2$ and that we only have {\sl one} (instead of two) parameter that labels the D-brane. However, inserting (\ref{regAdS2d-rho2-Sol}) into the $b^{-2}/2$-shift equation, we find that under no conditions can it be satisfied. We are thus led to the conjecture that this type of D-brane does not exist.

\subsection{\label{regAdS2d-rho1}Regular $AdS_2^{(d)}$ Branes - Gluing Map $\rho_1$}
\subsubsection{Shift Equations}
Again we choose the regular $u$-dependence, so that the gluing map $\rho_1$ fixes the one point function to be
\eq{\cor{\Theta_j(u\arrowvert z)}_{\alpha}=\brac{z-\bar{z}}^{-2h(j)}\brac{u-\bar{u}}^{2j}A_{\sigma}(j\arrowvert\alpha).}
The boundary two point function with degenerate field $t/2$, $t=1,b^{-2}$ is
\spliteq{G^{(2)}_{j,t,\alpha}(u_1,u_2\arrowvert z_1,z_2)=\brac{z_1-\bar{z}_1}&^{-2h(j)}\brac{z_2-\bar{z}_2}^{-2h(t/2)}\cdot\\
&\cdot \brac{u_1-\bar{u}_1}^{2j}\brac{u_2-\bar{u}_2}^{t}H^{(2)}_{j,t,\alpha}(u\arrowvert z),} 
with crossing ratios
\eq{z:=\frac{\abs{z_1-z_2}^2}{\brac{z_1-\bar{z}_1}\brac{z_2-\bar{z}_2}} \hskip .5cm {\rm and} \hskip .5cm
u:=-\frac{\abs{u_1-u_2}^2}{\brac{u_1-\bar{u}_1}\brac{u_2-\bar{u}_2}}.}
Solving the Knizhnik-Zamolodchikov equations for $t=1$ results in the same conformal blocks as for gluing map $\rho_2$, so they are given by equation (\ref{ConfBlocks-1/2}), again with parameters
\eq{a=-b^2(2j+2),\hskip .3cm b=-b^2(2j),\hskip .3cm c=-b^2(2j+1).}
The solution for $t=b^{-2}$ yields the conformal blocks (\ref{ConfBlocks}), again with parameters
\eq{\alpha=-2j,\hskip .3cm \beta=-b^{-2},\hskip .3cm \beta'=-2j-1-b^{-2},\hskip .3cm \gamma=-2j-b^{-2},}
just like for $\rho_2$.
The expansion coefficients are not altered here. They are simply
\eq{a^{j}_{\epsilon}(\alpha)=C_{\epsilon}(j)A_{\sigma}(j_{\epsilon}\arrowvert\alpha)}
for $t=1$ as well as $t=b^{-2}$.
In the limit $\abs{z_2-z_1}\rightarrow\infty$ followed by the same in the $u$'s, the same comments as in chapter \ref{regAdS2d-rho2} for $\rho_2$ apply. The shift equations that we produce read
\spliteq{\label{regAdS2d-rho1-Shift1}-\frac{1}{\pi}\Gamma(-b^2)\sin[2\pi b^2]\sin[\pi b^2(2j+1)]f_{\sigma}\brac{\frac{1}{2}}f_{\sigma}(j)&=\\
e^{-i\pi b^{2}j}\sin[\pi b^2(2j+2)]f_{\sigma}\brac{j+\frac{1}{2}}-e^{i\pi b^{2}j}&\sin[\pi b^2 2j]f_{\sigma}\brac{j-\frac{1}{2}}}
and
\spliteq{\label{regAdS2d-rho1-Shift2}\frac{e^{-i\pi(j+b^{-2})}}{\Gamma(1+b^2)}f_{\sigma}&\brac{\frac{b^{-2}}{2}}f_{\sigma}(j)=e^{-2\pi i\brac{j+\frac{b^{-2}}{2}}}f_{\sigma}\brac{j+\frac{b^{-2}}{2}}+\\
&+e^{2\pi i\brac{j+\frac{b^{-2}}{2}}}f_{-\sigma}\brac{j+\frac{b^{-2}}{2}}-e^{-i\pi b^{-2}}f_{\sigma}\brac{j-\frac{b^{-2}}{2}}.}

\subsubsection{Solving the Shift Equations}
Again we take a first step by solving the $1/2$-shift equation (\ref{regAdS2d-rho1-Shift1}) together with the reflection symmetry constraint (\ref{RefSymmCnstr-reg}). Here, we find the solution
\eq{\label{regAdS2d-rho1-Sol}f_{\sigma}(j\arrowvert m)=2\pi i\sigma\cos[\pi b^2/2]\frac{e^{-i\pi\frac{3b^2}{4}}}{\Gamma(-b^2)}e^{i\pi m}e^{-i\pi\sigma(m-\frac{1}{2})(2j+1)}\frac{\exp\ebrac{i\pi\frac{b^2}{4}(2j+1)^2}}{\sin[\pi b^2(2j+1)]},}
with $m\in\mathbb{Z}$. The same comments as in section \ref{regAdS2d-rho2} apply. Again, the $b^{-2}/2$-shift equation cannot be satisfied by this solution and we conjecture that this type of D-brane does also not exist.

\section{\label{Continuous}More on the Continuous D-Branes}
In this chapter we assemble our results (shift equations and solutions) concerning the continuous branes. The two point functions are always determined as shown in the corresponding sections of chapter \ref{Discrete} and thus, we do not write them down here again, but merely state our results. For some remarks on taking the limit $\Im(z_2)\rightarrow 0$ that is relevant here, see appendix \ref{ZeroLim}.  
The $1/2$-shift equations for the irregular $AdS_2^{(c)}$ brane with gluing maps $\rho_2$/$\rho_1$ have already been discussed in \cite{PST}/\cite{LOP}.

\subsection{\label{irrAdS2c-rho2-1}Irregular $AdS_2^{(c)}$ Branes - Gluing Maps $\rho_1$, $\rho_2$}
As before in the discrete case, we discover that the irregular continuous D-branes are isomorphic for gluing maps $\rho_1$, $\rho_2$. The shift equations are
\spliteq{\label{irrAdS2c-rho2-1-Shift1}\sigma\sqrt{\nu_b}\frac{\Gamma(-b^2)}{\Gamma(-2b^2)}C(1/2,0\arrowvert\alpha)\sin[\pi b^2(2j+1)]f_{\sigma}\brac{j}=&\\
\sin[\pi b^2(2j+2)]f_{\sigma}\brac{j+\frac{1}{2}}-&\sin[\pi b^2 2j]f_{\sigma}\brac{j-\frac{1}{2}}}
and
\spliteq{\label{irrAdS2c-rho2-1-Shift2}\nu_b^{\frac{b^{-2}}{2}}(1+b^2)e^{-i\pi\sigma\frac{b^{-2}}{2}}C(b^{-2}/2,0\arrowvert\alpha)f_{\sigma}(j)&=e^{-i\pi\frac{b^{-2}}{2}}f_{\sigma}\brac{j+\frac{b^{-2}}{2}}-\\
-e^{i\pi\brac{2j+\frac{b^{-2}}{2}}}f_{-\sigma}\brac{j+\frac{b^{-2}}{2}}+&e^{i\pi\frac{b^{-2}}{2}}f_{\sigma}\brac{j-\frac{b^{-2}}{2}}.}
In \cite{PST} and \cite{LOP}, the following solution to the $1/2$-shift equation (\ref{irrAdS2c-rho2-1-Shift1}) and the reflection symmetry constraint (\ref{RefSymmCnstr-irr}) has been proposed
\eq{\label{irrAdS2c-rho2-1-Sol}f_{\sigma}(j\arrowvert\alpha)=-\frac{\pi A_b}{\sqrt{\nu_b}}\frac{e^{-\alpha(2j+1)\sigma}}{\sin[\pi b^2(2j+1)]}.}
To obtain this solution, it was used that 
\eq{C(1/2,0\arrowvert\alpha)=-\frac{1}{\sqrt{\nu_b}}\frac{\Gamma(-2b^2)}{\Gamma(-b^2)}2\sinh(\alpha).}
Although we do not have an explicit expression for the bulk-boundary OPE coefficient $C(b^{-2}/2,0\arrowvert\alpha)$, we can still make a further reaching statement here. Plugging the solution (\ref{irrAdS2c-rho2-1-Sol}) into the $b^{-2}/2$-shift equation (\ref{irrAdS2c-rho2-1-Shift2}), we get
\spliteq{-\nu_b^{b^{-2}/2}&e^{-i\pi\sigma b^{-2}/2}(1+b^2)C(b^{-2}/2,0\arrowvert\alpha)=
e^{-i\pi b^{-2}/2}e^{-\alpha\sigma b^{-2}}+\\
&+e^{i\pi b^{-2}/2}e^{\alpha\sigma b^{-2}}-e^{i\pi b^{-2}/2}e^{\alpha\sigma b^{-2}}e^{2\pi ij}e^{2\alpha(2j+1)\sigma}.}
Since the L.H.S. is independent of $j$, so must be the R.H.S. The only way to ensure this and still get a physically meaningful result (i.e. $\alpha$ independent of $\sigma$ and $j$ as well as $j$ independent of $\sigma$), is to have 
\eq{\label{Rstr-irrAdS2c-rho2-1}\alpha=i\frac{\pi}{2}q,\hskip .1cm q\in\mathbb{Q}\hskip .3cm \mathrm{and} \hskip .3cm j\in j_0+\mathrm{LCM}\brac{\frac{1}{1-q},\frac{1}{1+q}}\mathbb{Z},}
where $j_0\in\mathbb{R}$ is a fixed offset and 'LCM' denotes the least common multiple. The restriciton on $\alpha$ stems essentially from requiring the LCM to exist. The meaning of this result is twofold: Firstly, the irregular $AdS_2^{(c)}$ branes are actually not labelled by a continuous parameter, but only a discrete series of $\alpha$'s gives consistent boundary states. Secondly, we find that strings in the physical spectrum $j\in -1/2+i\mathbb{R}_+$ do not couple to these branes.

\subsection{\label{regAdS2c-rho2}Regular $AdS_2^{(c)}$ Branes - Gluing Map $\rho_2$}
We have the following shift equations
\spliteq{\label{regAdS2c-rho2-Shift1}\sigma\sqrt{\nu_b}\frac{\Gamma(-b^2)}{\Gamma(-2b^2)}C(1/2,0\arrowvert\alpha)\sin[\pi b^2(2j+1)]f_{\sigma}\brac{j}=&\\
e^{-i\pi b^{2}j}\sin[\pi b^2(2j+2)]f_{\sigma}\brac{j+\frac{1}{2}}+e^{i\pi b^{2}j}&\sin[\pi b^2 2j]f_{\sigma}\brac{j-\frac{1}{2}}}
and
\spliteq{\label{regAdS2c-rho2-Shift2}\nu_b^{\frac{b^{-2}}{2}}(1+b^2)&e^{-i\pi\sigma\frac{b^{-2}}{2}}e^{-i\pi(j+\frac{b^{-2}}{2})}C(b^{-2}/2,0\arrowvert\alpha)f_{\sigma}(j)=\\
e^{-2\pi i\brac{j+\frac{b^{-2}}{2}}}f_{\sigma}&\brac{j+\frac{b^{-2}}{2}}-e^{4\pi i\brac{j+\frac{b^{-2}}{2}}}f_{-\sigma}\brac{j+\frac{b^{-2}}{2}}-f_{\sigma}\brac{j-\frac{b^{-2}}{2}}.}
Again we try to solve the $1/2$-shift equation together with constraint (\ref{RefSymmCnstr-reg}) first. The following one parameter set of solutions can be achieved
\eq{\label{regAdS2c-rho2-Sol}f^{(m)}_{\sigma}(j\arrowvert\alpha_m)=\sigma e^{i\pi\frac{b^2}{4}(2j+1)^2}e^{i\frac{\pi}{2}(2m+1)(2j+1)\sigma},}
for $m\in\mathbb{Z}$ and $\alpha_m$ solving
\eq{\label{Rstr1-regAdS2c-rho2}\sinh(\alpha_m)=i(-)^{m+1}e^{i\pi\frac{b^2}{4}}\sin(\pi b^2/2).}
Thus, we find a restriction on the possibly continuous D-branes already in this step - they are labelled by a discrete parameter $\alpha_m$. Note that equations (\ref{regAdS2c-rho2-Sol}) and (\ref{Rstr1-regAdS2c-rho2}) really give a different solution for each $m\in\mathbb{Z}$ we choose. Plugging this solution into the $b^{-2}/2$-shift equation, we obtain a further restriction on $j$, namely
\eq{\label{Rstr2-regAdS2c-rho2}j\in j_0 +\mathrm{LCM}\brac{\frac{1}{2-2m},\frac{1}{4+2m}}\mathbb{Z}.}

\subsection{\label{regAdS2c-rho1}Regular $AdS_2^{(c)}$ Branes - Gluing Map $\rho_1$}
For the shift equations, we obtain
\spliteq{\label{regAdS2c-rho1-Shift1}\sigma\sqrt{\nu_b}\frac{\Gamma(-b^2)}{\Gamma(-2b^2)}C(1/2,0\arrowvert\alpha)\sin[\pi b^2(2j+1)]f_{\sigma}\brac{j}&=\\
e^{-i\pi b^{2}j}\sin[\pi b^2(2j+2)]f_{\sigma}\brac{j+\frac{1}{2}}-e^{i\pi b^{2}j}&\sin[\pi b^2 2j]f_{\sigma}\brac{j-\frac{1}{2}}}
and
\spliteq{\label{regAdS2c-rho1-Shift2}\nu_b^{\frac{b^{-2}}{2}}(1+b^2)e^{-i\pi\sigma\frac{b^{-2}}{2}}e^{-i\pi(j-\frac{b^{-2}}{2})}C(b^{-2}/2,0\arrowvert\alpha)&f_{\sigma}(j)=\\
e^{-2\pi i j}f_{\sigma}\brac{j+\frac{b^{-2}}{2}}+e^{2\pi i(j+b^{-2})}f_{-\sigma}\brac{j+\frac{b^{-2}}{2}}&-e^{-i\pi b^{-2}}f_{\sigma}\brac{j-\frac{b^{-2}}{2}}.}
The $1/2$-shift equation together with constraint (\ref{RefSymmCnstr-reg}) can be solved analogously to the former case. We get
\eq{\label{regAdS2c-rho1-Sol}f^{(m)}_{\sigma}(j\arrowvert\alpha_m)=\sigma e^{i\pi\frac{b^2}{4}(2j+1)^2}e^{i\frac{\pi}{2}(2m+1)(2j+1)\sigma},}
with $m\in\mathbb{Z}$ and this time $\alpha_m$ a solution of
\eq{\label{Rstr1-regAdS2c-rho1}\sinh(\alpha_m)=i(-)^{m+1}e^{i\pi\frac{b^2}{4}}\cos(\pi b^2/2).}
Furthermore, the $b^{-2}/2$-shift equation enforces a condition on the $j$'s, which is
\eq{\label{Rstr2-regAdS2c-rho1}j\in j_0+\mathrm{LCM}\brac{\frac{1}{1-2m},\frac{1}{3+2m}}\mathbb{Z}.}

\section{\label{Conclusion}Conclusions and Outlook}
We have shown that the boundary $H_3^+$ model possesses a variety of D-brane types, regular and irregular, discrete and continuous, that must all be analysed, case by case, and checked for consistency. We initiated the systematic derivation of $1/2$- and $b^{-2}/2$-shift equations, giving all of them for gluing maps $\rho_1$ and $\rho_2$ (see chapter \ref{GlueConds} for a definition of the gluing maps). Our achievements are summarised in Table \ref{T2} below.\\
We also discussed possible solutions and showed that the irregular $AdS_2^{(d)}$ branes of \cite{SR1} only couple to strings in finite dimensional $SL(2)$ representations $j\in \mathbb{Z}/2$. As the physical spectrum of the bulk $H_3^+$ theory consists of strings in the infinite dimensional continuous representations $j\in -1/2+i\mathbb{R}_+$, this would mean that such branes can be discarded, as no physical states couple to them. In addition, we argued that also the irregular $AdS_2^{(c)}$ branes of \cite{PST} and \cite{LOP} do not couple consistently to the physical $H_3^+$ strings. Moreover, we found that their couplings to closed strings can only be consistent if the D-brane label $\alpha\in i\frac{\pi}{2}\mathbb{Q}$. This means that the non-trivial (i.e. interacting with closed strings) irregular
\begin{table}
\begin{tabular}{|c||c|c|c|c|c|} \hline
             &$u$-dependence   
                       &\multicolumn{2}{c|}{shift equation (continuous)}   
                                 &\multicolumn{2}{c|}{shift equation (discrete)}\\
             &         
                       &for $\Theta_{1/2}?$    &for $\Theta_{b^{-2}/2}?$ 
                                 &for $\Theta_{1/2}?$   &for $\Theta_{b^{-2}/2}?$\\ \hline\hline
$\rho_1$     &$|u-\bar{u}|^{2j}$ 
                       &\cite{LOP}/(\ref{irrAdS2c-rho2-1-Shift1})                    
                                               &(\ref{irrAdS2c-rho2-1-Shift2})
                                 &(\ref{irrAdS2d-rho1-Shift1})   
                                                        &(\ref{irrAdS2d-rho1-Shift2})\\ \cline{2-6}
             &$(u-\bar{u})^{2j}$ 
                       &\cite{GKS}/(\ref{regAdS2c-rho1-Shift1})     
                                               &(\ref{regAdS2c-rho1-Shift2})
                                 &\cite{GKS}/(\ref{regAdS2d-rho1-Shift1})                       
                                                        &\cite{GKS}/(\ref{regAdS2d-rho1-Shift2})\\ \hline
$\rho_2$     &$|u+\bar{u}|^{2j}$ 
                       &\cite{PST}
                                               &(\ref{irrAdS2c-rho2-1-Shift2})                 
                                 &\cite{SR1}                         
                                                        &(\ref{irrAdS2d-rho2-Shift2})\\ \cline{2-6}
             &$(u+\bar{u})^{2j}$ 
                       &(\ref{regAdS2c-rho2-Shift1})                    
                                               &(\ref{regAdS2c-rho2-Shift2})
                                 &(\ref{regAdS2d-rho2-Shift1})                                        
                                                        &(\ref{regAdS2d-rho2-Shift2})\\ \hline
$\rho_3$     &$|-1+u\bar{u}|^{2j}$ 
                       &---                    &---                   
                                 &---                   &---\\ \cline{2-6}  
             &$(-1+u\bar{u})^{2j}$       
                       &---                    &---
                                 &\cite{GKS}            &\cite{GKS}\\ \hline
$\rho_4$     &$(1+u\bar{u})^{2j}$       
                       &---                    &---
                                 &\cite{PST}    &---\\ \hline                      
\end{tabular}
\caption{\label{T2} Classes of D-brane solutions - the table as it presents itself now. In the cases of discrete and continuous regular $\rho_1$, we find shift equations that are slightly modified from those given in \cite{GKS}. However, recall that this slight modification has great impact on the solvability of the equations, as discussed in chapters \ref{regAdS2d-rho1} and \ref{regAdS2c-rho1}.}
\end{table}
$AdS_2^{(c)}$ branes are in fact {\sl not} labelled by a continuous parameter. Also, only a very restricted set of fields with labels $j=j_n(\alpha)$ as in equation (\ref{Rstr-irrAdS2c-rho2-1}), which transform in certain discrete $SL(2)$ representations, can couple to such a D-brane. The nonexistence (or rather inconsistency) of $AdS_2^{(c)}$ branes had actually been conjectured in \cite{BP1} and \cite{BP2}. Our conclusion is, that the irregular $AdS_2^{(c)}$ of \cite{PST} and \cite{LOP} as well as the irregular $AdS_2^{(d)}$ branes of \cite{SR1} can be discarded from the boundary $H_3^+$ theory. Nevertheless can these D-branes still turn out to be important in view of string theory on $AdS_3$ or the cigar CFT (string in euclidean black hole background), because the physical spectrum of these theories is richer (see \cite{MaldaOoguri1} and \cite{DVV}, respectively). Additionally, there might be subtleties in going from euclidean to lorentzian $AdS_3$. This has been mentioned for example in \cite{CD1}. It must also be emphasised that it remains unclear, whether or not different solutions (other than (\ref{irrAdS2d-rho2-Sol}) and (\ref{irrAdS2c-rho2-1-Sol})) to our systems of shift equations (\ref{irrAdS2d-rho2-Shift2}), (\ref{irrAdS2d-rho2-Shift1}) and (\ref{irrAdS2c-rho2-1-Shift1}), (\ref{irrAdS2c-rho2-1-Shift2}) (both together with the constraint (\ref{RefSymmCnstr-irr})) can be constructed. A careful analysis of these questions is still in need.\\
Furthermore, based on our attempts to solve the shift equations for the regular discrete D-branes with gluing maps $\rho_1$, $\rho_2$, we would have to conclude that these D-brane types do not exist at all. Again, it would be very valuable to have rigorous proofs that there are no solutions to our systems of shift eqations (\ref{regAdS2d-rho2-Shift1}), (\ref{regAdS2d-rho2-Shift2}) and (\ref{regAdS2d-rho1-Shift1}), (\ref{regAdS2d-rho1-Shift2}) (in both cases together with the constraint (\ref{RefSymmCnstr-reg})).\\
Concerning the regular $AdS_2^{(c)}$ branes, we have found a one parameter ($m\in\mathbb{Z}$) set of possible solutions, (\ref{regAdS2c-rho2-Sol}) and (\ref{regAdS2c-rho1-Sol}) respectively. Once more, we cannot exclude the possibility that other solutions might exist. The solutions that we gave are basically as restricted as the irregular ones: Non-trivial branes are labelled by a discrete parameter $\alpha_m$ (given by equations (\ref{Rstr1-regAdS2c-rho2}) and (\ref{Rstr1-regAdS2c-rho1}) respectively). Only very specific fields with labels $j=j_n(m)$ as in (\ref{Rstr2-regAdS2c-rho2}) and (\ref{Rstr2-regAdS2c-rho1}) respectively, do couple consistently. Let us recall here, that our restrictions (\ref{Rstr-irrAdS2c-rho2-1}), (\ref{Rstr2-regAdS2c-rho2}) and (\ref{Rstr2-regAdS2c-rho1}) on the continuous D-branes are all necessary conditions. The solutions (\ref{regAdS2c-rho2-Sol}) and (\ref{regAdS2c-rho1-Sol}) might be even more restricted or turn out to be entirely inconsistent. To decide this question, one would have to use the explicit form of the bulk-boundary OPE coefficient $C(b^{-2}/2,0\arrowvert\alpha)$.\\
Another point we like to stress is the following: For the discrete D-branes as well as for the continuous ones, the
qualitative behaviour of the regular solutions that we have found is always strikingly different from the irregular solutions. It would be interesting to understand what exactly happens when passing from the regular to the irregular dependence.\\
Let us also remark that there is a different approach to boundary $H_3^+$, namely via the direct construction of boundary states. As this approach is equivalent to the one we have used here, it is natural to ask how our results translate into the boundary state formalism.\\
Looking at Table \ref{T2}, we have to admit that it is still incomplete. In this paper, we have only given the shift equations for gluing maps $\rho_1$, $\rho_2$. The case of gluing conditions $\rho_3$ and $\rho_4$ is under investigation and will appear in a follow-up publication.\\ 
But what is achieved after having filled the table completely? We will then have a good overview over those D-branes that are certainly not and those who {\sl might} be consistent boundary states of the $H_3^+$ boundary theory. It is important to note that those D-branes that are not ruled out by our systematic procedure, are not guaranteed to be consistent boundary states. In order to achieve certainty in that matter, the boundary three point functions of open strings ending on such D-branes have to be studied and shown to be fully consistent.\footnote{Such correlators have been given recently for the case of $AdS_2$ boundary conditions in \cite{HoRi}.} We are going to tackle this problem in the near future.\\
\vskip 2mm \noindent 
{\bf Acknowledgements:} We would like to thank J\"org Teschner for some helpful comments in the very early stages of the project. H.A. acknowledges financial support by the DFG-Graduiertenkolleg No. 282. The work of M.F. is partially supported by the European Union network HPRN-CT-2002-00325 (EUCLID).
\newpage

\appendix
\section{Exact Two Point Function Involving $\Theta_{b^{-2}/2}$}
\subsection{\label{KniZa}Solution of the Knizhnik-Zamolodchikov Equation}
In (\ref{B2pt}) we have given the general form of the two point function $G^{(2)}_{j,\alpha}(u_i\arrowvert z_i)$ fixed by the Ward identities. To this we apply the Knizhnik-Zamolodchikov equation for $z_2$ which reads
\spliteq{-\frac{1}{b^2}\partial_{z_2}G^{(2)}_{j,\alpha}(u_i\arrowvert z_i)&=\\
\sum_a {\cal D}_{b^{-2}/2}^a(u_2) \otimes &\ebrac{\frac{{\cal D}_j^a(u_1)}{z_2-z_1}+\frac{\rho\left(\bar{{\cal D}}_j^a(\bar{u}_1)\right)}{z_2-\bar{z}_1}+\frac{\rho\left(\bar{{\cal D}}_{b^{-2}/2}^a(\bar{u}_2)\right)}{z_2-\bar{z}_2}}G^{(2)}_{j,\alpha}(u_i\arrowvert z_i).}
Mapping $z_1\rightarrow 0$, $\bar{z}_2\rightarrow 1$ and $\bar{z}_1\rightarrow\infty$ (i.e. $z_2\rightarrow z$) brings this equation to standard form
\spliteq{-&b^{-2}z(z-1)\partial_{z}H^{(2)}_{j,\alpha}(u\arrowvert z)=u(u-1)(u-z)\partial^2_u H^{(2)}_{j,\alpha}+\\
&+\set{\ebrac{1-2b^{-2}}u^2+\ebrac{b^{-2}-2j-2}uz+\ebrac{2j+b^{-2}}u+z}\partial_u H^{(2)}_{j,\alpha}+\\
&+\set{b^{-4}u+\ebrac{b^{-2}j-b^{-4}/2}z-b^{-2}j}H^{(2)}_{j,\alpha}.}
It is solved by (see \cite{JT1}) $H^{(2)}_{j,\alpha}=\sum_{\epsilon=+,-,x} a^{j}_{\epsilon}(\alpha){\cal F}^{s}_{j,\epsilon}$ with
\spliteq{\label{ConfBlocks}{\cal F}^{s}_{j,+}(u\arrowvert z)&=z^{-j}(1-z)^{-b^{-2}/2}F_1(\alpha,\beta,\beta';\gamma\arrowvert u;z),\\
{\cal F}^{s}_{j,-}(u\arrowvert z)&=z^{\beta-\gamma+1-j}(1-z)^{\gamma-\alpha-1-b^{-2}/2}(u-z)^{-\beta}\cdot\\
&\cdot F_1\left(1-\beta',\beta,\alpha+1-\gamma;2+\beta-\gamma\left\arrowvert\frac{z}{z-u};\frac{z}{z-1}\right)\right.,\\
{\cal F}^{s}_{j,x}(u\arrowvert z)&=z^{-j}(1-z)^{-b^{-2}/2}e^{i\pi(\alpha+1-\gamma)}\frac{\Gamma(\alpha)\Gamma(\gamma-\beta)}{\Gamma(\alpha+1-\beta)\Gamma(\gamma-1)}\cdot\\
&\cdot\left\{u^{-\alpha}F_1\left(\alpha,\alpha+1-\gamma,\beta';\alpha+1-\beta\left\arrowvert\frac{1}{u};\frac{z}{u}\right)\right.-\right.\\
&\left. -e^{-i\pi\alpha}\frac{\Gamma(\alpha+1-\beta)\Gamma(1-\gamma)}{\Gamma(\alpha+1-\gamma)\Gamma(1-\beta)}F_1(\alpha,\beta,\beta';\gamma\arrowvert u;z)\right\}.}
The function $F_1(\alpha,\beta,\beta',\gamma\arrowvert u;z)$ is a generalized hypergeometric function, namely the first one of Appell's double hypergeometric functions (see the book \cite{Exton} for more information). For the parameters we find
\eq{\alpha=\beta=-b^{-2},\hskip .3cm \beta'=-2j-1-b^{-2},\hskip .3cm \gamma=-2j-b^{-2}.}

\subsection{\label{LinComb}Finding the Correct Linear Combination of Conformal Blocks}
In order to obtain the exact result for the boundary two point function (\ref{B2pt}), all that is left to do is determine the coefficients $a^{j}_{\epsilon}(\alpha)$, i.e. find the correct linear combination of conformal blocks (\ref{ConfBlocks}). To this end, we use the operator product expansion (OPE) on the L.H.S. of (\ref{B2pt}) to obtain
\spliteq{&G^{(2)}_{j,\alpha}(u_1,u_2\arrowvert z_1,z_2)\simeq\\
&\simeq\abs{z_2-z_1}^{-2j}\abs{z_1-\bar{z}_1}^{-2h(j_+)}\abs{u_1+\bar{u}_1}^{2j+b^{-2}}C_+(j)A_{\sigma}(j_+\arrowvert\alpha)+\\
&+\abs{z_2-z_1}^{2j+2}\abs{u_2-u_1}^{2b^{-2}}\abs{z_1-\bar{z}_1}^{-2h(j_-)}\abs{u_1+\bar{u}_1}^{2j-b^{-2}}C_-(j)A_{\sigma}(j_-\arrowvert\alpha)+\\
&+\abs{z_2-z_1}^{-2j}\abs{u_2-u_1}^{2(2j+1+b^{-2})}\abs{z_1-\bar{z}_1}^{-2h(j_x)}\abs{u_1+\bar{u}_1}^{-2j-2-b^{-2}}\cdot\\
&\cdot C_x(j)A_{\sigma}(j_x\arrowvert\alpha).}
On the R.H.S. we can also take the limit $\abs{z_2-z_1}\rightarrow 0$ ($\Rightarrow z\rightarrow 0+$) followed by $\abs{u_2-u_1}\rightarrow 0$ ($\Rightarrow u\rightarrow 0+$). We have to be careful to take appropriate analytic continuations of Appell's function $F_1$ (see \cite{Exton} or Appendix \ref{Appell}) and the Hypergeometric Function (see Appendix \ref{Hypergeo}). It behaves as follows
\spliteq{&F_1(\alpha,\beta,\beta',\gamma\arrowvert u;z)\simeq 1,\\
&F_1\left(1-\beta',\beta,\alpha+1-\gamma,2+\beta-\gamma\left\arrowvert\frac{z}{z-u};\frac{z}{z-1}\right)\right.\simeq 1,\\
&F_1\left(\alpha,\alpha+1-\gamma,\beta',\alpha+1-\beta\left\arrowvert\frac{1}{u};\frac{z}{u}\right)\right.\simeq\\
&\simeq\frac{\Gamma(2j+1+b^{-2})}{\Gamma(2j+1)\Gamma(1+b^{-2})}e^{i\pi b^{-2}}u^{-b^{-2}}+\frac{\Gamma(-2j-1-b^{-2})}{\Gamma(-b^{-2})\Gamma(-2j)}e^{-i\pi (2j+1)}u^{2j+1}.}
Using these formulae when expanding the conformal blocks (\ref{ConfBlocks}) and comparing to the OPE expansion, we find precisely that
\eq{a^{j}_{\epsilon}(\alpha)=C_{\epsilon}(j)A_{\sigma}(j_{\epsilon}\arrowvert\alpha).}
This concludes our computation of the exact boundary two point function (\ref{B2pt}).

\section{Two Different Limits of the Exact Boundary Two Point Function}
\subsection{\label{InftyLim}$\abs{z_2-z_1}\rightarrow\infty$ followed by $\abs{u_2-u_1}\rightarrow\infty$}
We start from 
\spliteq{\label{ExB2pt}G^{(2)}_{j,\alpha}(u_1,u_2&\arrowvert z_1,z_2)=\abs{z_1-\bar{z}_1}^{2[h(b^{-2}/2)-h(j)]}\abs{z_1-\bar{z}_2}^{-4h(b^{-2}/2)}\cdot\\
&\cdot\abs{u_1+\bar{u}_1}^{2j-b^{-2}}\abs{u_1+\bar{u}_2}^{2b^{-2}}\sum_{\epsilon=+,-,x}C_{\epsilon}(j)A_{\sigma}(j_{\epsilon}\arrowvert\alpha){\cal F}^{s}_{j,\epsilon}(u\arrowvert z),}
with conformal blocks ${\cal F}^{s}_{j,\epsilon}(u\arrowvert z)$ given in (\ref{ConfBlocks}). Taking 
$\abs{z_2-z_1}\rightarrow\infty$ implies $z\rightarrow 1-$ (analogous result in the $u$'s). In this limit, we obtain the following asymptotic behaviour of the conformal blocks (carefully using the appropriate analytic continuations of Appell's function $F_1$; see Appendix \ref{Appell} or \cite{Exton}):
\spliteq{{\cal F}^{s}_{j,+}(u\arrowvert z)&\simeq (1-u)^{b^{-2}}(1-z)^{1+b^{-2}/2}\cdot\frac{\Gamma(-2j-b^{-2})\Gamma(-1-b^{-2})}{\Gamma(-b^{-2})\Gamma(-2j-1-b^{-2})},\\
{\cal F}^{s}_{j,-}(u\arrowvert z)&\simeq e^{i\pi b^{-2}}(1-u)^{b^{-2}}(1-z)^{1+b^{-2}/2}\cdot\frac{\Gamma(2j+2)\Gamma(-1-b^{-2})}{\Gamma(-b^{-2})\Gamma(2j+1)},\\
{\cal F}^{s}_{j,x}(u\arrowvert z)&\simeq e^{i\pi(2j+b^{-2})}(1-u)^{b^{-2}}(1-z)^{1+b^{-2}/2}\cdot\\
&\cdot\frac{\Gamma(-2j)\Gamma(2j+b^{-2}+1)\Gamma(-2j-b^{-2})\Gamma(-1-b^{-2})}{\Gamma(-2j-b^{-2}-1)\Gamma(2j+1)\Gamma(1+b^{-2})\Gamma(-2j-b^{-2}-1)}.}
Employing the redefinition (\ref{ReDef}) $f_{\sigma}(j):=\nu_b^j\Gamma(1+b^2(2j+1))A_{\sigma}(j\arrowvert\alpha)$, we then find 
\spliteq{C_+(j){\cal F}^{s}_{j,+}(u\arrowvert z)A_{\sigma}(j_+\arrowvert\alpha)\simeq -(1-u)&^{b^{-2}}(1-z)^{1+b^{-2}/2}\cdot\\
\cdot b^{-2}\nu_b^{-j-b^{-2}/2}&\frac{\Gamma(-1-b^{-2})}{\Gamma(-b^{-2})}\frac{f_{\sigma}(j_+)}{\Gamma(1+b^2(2j+1))},\\
C_-(j){\cal F}^{s}_{j,-}(u\arrowvert z)A_{\sigma}(j_-\arrowvert\alpha)\simeq -(1-u)&^{b^{-2}}(1-z)^{1+b^{-2}/2}\cdot\\
\cdot e^{i\pi b^{-2}}b^{-2}\nu_b^{-j-b^{-2}/2}&\frac{\Gamma(-1-b^{-2})}{\Gamma(-b^{-2})}\frac{f_{\sigma}(j_-)}{\Gamma(1+b^2(2j+1))},\\
C_x(j){\cal F}^{s}_{j,x}(u\arrowvert z)A_{\sigma}(j_x\arrowvert\alpha)\simeq -(1-u)&^{b^{-2}}(1-z)^{1+b^{-2}/2}\cdot\\
\cdot e^{i\pi 2j+b^{-2}}b^{-2}\nu_b^{-j-b^{-2}/2}&\frac{\Gamma(-1-b^{-2})}{\Gamma(-b^{-2})}\frac{f_{\sigma}(j_x)}{\Gamma(1+b^2(2j+1))}.}
With the help of 
\eq{1-z=\frac{4\Im(z_1)\Im(z_2)}{\abs{z_2-\bar{z}_1}^2},\hskip 1cm 1-u=\frac{4\Re(u_1)\Re(u_2)}{\abs{u_2+\bar{u}_1}^2},}
it is easy to check, that, together with the factors in (\ref{ExB2pt}), this gives the correct $u$- and $z$-dependence expected of a product of one point functions. Assembling terms from L.H.S. and R.H.S. finally gives us the desired shift equation
\eq{\frac{f_{\sigma}(b^{-2}/2)f_{\sigma}(j)}{\Gamma(1+b^2)}=f_{\sigma}(j_+)-e^{2\pi ij_+}f_{-\sigma}(j_+)+e^{i\pi b^{-2}}f_{\sigma}(j_-).}

\subsection{\label{ZeroLim}$\Im(z_2)\rightarrow 0$}
For $\Im(z_2)\rightarrow 0$, the crossing ratio $z=\frac{\abs{z_2-z_1}^2}{\abs{z_2-\bar{z}_1}^2}\rightarrow 1-$, i.e. the limit on the R.H.S. of the two point function is the same as above in section \ref{InftyLim}. What changes is the L.H.S., where we now use the bulk-boundary OPE of $\Theta_{b^{-2}/2}$:
\spliteq{\Theta_{b^{-2}/2}(u_2\arrowvert z_2)&\simeq (z_2-\bar{z}_2)^{1-b^{-2}/2}(u_2+\bar{u}_2)^{b^{-2}}C(b^{-2}/2,0\arrowvert\alpha)+\dots\\
&=\abs{z_2-\bar{z}_2}^{1-b^{-2}/2}\abs{u_2+\bar{u}_2}^{b^{-2}}(\sigma)^{b^{-2}}C(b^{-2}/2,0\arrowvert\alpha)+\dots,}
with bulk-boundary OPE coefficient $C(b^{-2}/2,0\arrowvert\alpha)$. Therefore, we obtain on the L.H.S.
\spliteq{G^{(2)}_{j,\alpha}(u_1,u_2\arrowvert z_1,z_2)\simeq\abs{z_1-\bar{z}_1}^{-2h(j)}&\abs{z_2-\bar{z}_2}^{-2h(b^{-2}/2)}\abs{u_1+\bar{u}_1}^{2j}\abs{u_2+\bar{u}_2}^{b^{-2}}\cdot\\
&\cdot e^{i\pi(1-\sigma)\frac{b^{-2}}{2}}C(b^{-2}/2,0\arrowvert\alpha)f_{\sigma}(j)+\dots.}
Equating with the R.H.S. gives our shift equation (\ref{irrAdS2c-rho2-1-Shift2}).

\section{OPE Coefficients}
We obtain the bulk OPE coefficients from the structure constants that were given in \cite{JT2}. We only need to be careful about the different normalisations of field operators. In \cite{JT2}, the operators $\phi_j(u|z)$ are used, whereas here (as well as in \cite{PST}) we are working with $\Theta_j(u|z):=B^{-1}(j)\phi_j(u|z)$, where $B(j)=(2j+1)R(j)/\pi$, $R(j)$ being the reflection amplitude, see (\ref{Rj}). With this, the structure constants $D(j,j_1,j_2)$ of \cite{JT2} have to be "dressed" by some factors of $B^{-1}$:
\eq{C(j,j_1,j_2):=D(j,j_1,j_2)B^{-1}(j_1)B^{-1}(j_2).}

\subsection{\label{OPE1}Bulk OPE Coefficients for the OPE with $\Theta_{1/2}$}
For completeness we give the bulk OPE coefficients with the degenerate field $\Theta_{1/2}$, although they are also written in \cite{PST}, using the same normalisation as we do. Since $\Theta_{1/2}$ is degenerate, the OPE is highly restricted. Only the field operators with $j_+=j+1/2$ and $j_-=j-1/2$ do occur. The corresponding coefficients are
\eq{C_+(j)=1, \hskip .5cm C_-(j)=\frac{1}{v_b}\frac{\Gamma(-b^2(2j+1))\Gamma(1+2b^2j)}{\Gamma(1+b^2(2j+1))\Gamma(-2b^2j)}.}

\subsection{\label{OPE2}Bulk OPE Coefficients for the OPE with $\Theta_{b^{-2}/2}$}
The singular vector labelled by $b^{-2}/2$ restricts the possibly occuring field operators in the operator product to those with labels $j_+:=j+b^{-2}/2$, $j_-:=j-b^{-2}/2$, $j_x:=-j-1-b^{-2}/2$. The corresponding OPE coefficients can be easily calculated. We obtain
\spliteq{C_+(j)=1&, \hskip .5cm C_-(j)=-\nu_b^{-b^{-2}}\ebrac{b^2(2j+1)}^{-2},\\ 
C_x(j)=-\frac{\nu_b^{-2j-1-b^{-2}}}{b^4}&\frac{\Gamma(1+b^{-2})}{\Gamma(1-b^{-2})}\frac{\Gamma(1+2j)\Gamma(-1-2j-b^{-2})\Gamma(-b^2(2j+1))}{\Gamma(-2j)\Gamma(2+2j+b^{-2})\Gamma(1+b^2(2j+1))}.}

\section{\label{RefCnstr}A Further Constraint on the One Point Amplitude from Reflection Symmetry}

\subsection{The Irregular One Point Amplitudes}
Due to the reflection symmetry (\ref{RefSymm}), the one point amplitude has to obey
\spliteq{\frac{\pi}{2j+1}&\abs{u\mp\bar{u}}^{2j}A_{\sigma}(j\arrowvert\alpha)=\\
&=-R(-j-1)\int_{\mathbb{C}}d^2u'\abs{u-u'}^{4j}\abs{u'\mp\bar{u}'}^{-2j-2}A_{\sigma'}(-j-1\arrowvert\alpha).}
The upper sign corresponds to gluing map $\rho_1$, the lower sign to $\rho_2$. Note that $\sigma '\equiv\sigma(u')$. Since we can always expand $A_{\sigma '}(-j-1\arrowvert\alpha)=A^0(-j-1\arrowvert\alpha)+\sigma 'A^1(-j-1\arrowvert\alpha)$, we need to compute the integrals ($\epsilon\in\set{0,1}$):
\eq{I^{\mp}_{\epsilon}:=\int_{\mathbb{C}}d^2u'\abs{u-u'}^{4j}\abs{u'\mp\bar{u}'}^{-2j-2}(\sigma ')^{\epsilon}.}

\subsubsection{Gluing Map $\rho_1$ - Calculation of $I^{-}_{\epsilon}$}
Assume $u_2>0$. We split the integral into
\spliteq{I^{-}_{\epsilon}&=(-)^{\epsilon}\int_{-\infty}^{+\infty}du'_1\int_{-\infty}^{0}du'_2\ebrac{(u_1-u'_1)^2+(u_2-u'_2)^2}^{2j}(-2u'_2)^{-2j-2}+\\
&+\int_{-\infty}^{+\infty}du'_1\int_{0}^{u_2}du'_2\ebrac{(u_1-u'_1)^2+(u_2-u'_2)^2}^{2j}(2u'_2)^{-2j-2}+\\
&+\int_{-\infty}^{+\infty}du'_1\int_{u_2}^{+\infty}du'_2\ebrac{(u_1-u'_1)^2+(u_2-u'_2)^2}^{2j}(2u'_2)^{-2j-2}\\
&\equiv (-)^{\epsilon}I_1^{>}+I_2^{>}+I_3^{>}.}
Being careful about signs and using some Gamma function identities (see \ref{Gamma}), we obtain
\eq{I_1^{>}=-\frac{\pi}{2j+1}\abs{u-\bar{u}}^{2j},\hskip .5cm I_2^{>}=-I_3^{>}.}
Now, assume $u_2<0$. In this case, we choose the following splitting
\spliteq{I^{-}_{\epsilon}&=(-)^{\epsilon}\int_{-\infty}^{+\infty}du'_1\int_{-\infty}^{u_2}du'_2\ebrac{(u_1-u'_1)^2+(u_2-u'_2)^2}^{2j}(-2u'_2)^{-2j-2}+\\
&+(-)^{\epsilon}\int_{-\infty}^{+\infty}du'_1\int_{u_2}^{0}du'_2\ebrac{(u_1-u'_1)^2+(u_2-u'_2)^2}^{2j}(-2u'_2)^{-2j-2}+\\
&+\int_{-\infty}^{+\infty}du'_1\int_{0}^{+\infty}du'_2\ebrac{(u_1-u'_1)^2+(u_2-u'_2)^2}^{2j}(2u'_2)^{-2j-2}\\
&\equiv (-)^{\epsilon}I_1^{<}+(-)^{\epsilon}I_2^{<}+I_3^{<}.}
This time we get
\eq{I_1^{<}=-I_2^{<},\hskip .5cm I_3^{<}=-\frac{\pi}{2j+1}\abs{u-\bar{u}}^{2j}.}
Assembling, we obtain
\eq{I^{-}_{\epsilon}=-\frac{\pi}{2j+1}\abs{u-\bar{u}}^{2j}(-\sigma)^{\epsilon}.}

\subsubsection{Gluing Map $\rho_2$ - Calculation of $I^{+}_{\epsilon}$}
Splitting the integral as before and renaming the integration variables, it is easy to see that
\eq{I^{+}_{\epsilon}=I^{-}_{\epsilon}(u_1\leftrightarrow u_2)=-\frac{\pi}{2j+1}\abs{u+\bar{u}}^{2j}(-\sigma)^{\epsilon}.}

\subsubsection{The Constraint for the Irregular One Point Amplitudes}
Putting things together, we arrive at the constraint
\eq{A_{\sigma}(j\arrowvert\alpha)=R(-j-1)A_{-\sigma}(-j-1\arrowvert\alpha).}
Using the definition of the reflection amplitude (\ref{Rj}), we are led to redefine the one point amplitude
\eq{f_{\sigma}(j):=\nu _b^j\Gamma(1+b^2(2j+1))A_{\sigma}(j\arrowvert\alpha)}
(note that we have dropped the $\alpha$-dependence of $f_{\sigma}$). For this redefined one point amplitude, the constraint simply reads
\eq{\label{RefSymmCnstr-irr}f_{\sigma}(j)=-f_{-\sigma}(-j-1).}

\subsection{The Regular One Point Amplitudes}
This time we need to compute the integrals ($\epsilon\in\set{0,1}$):
\eq{I^{\mp}_{\epsilon}:=\int_{\mathbb{C}}d^2u'\abs{u-u'}^{4j}\brac{u'\mp\bar{u}'}^{-2j-2}(\sigma ')^{\epsilon}.} Up to a sign, the result is very much the same as before:
\eq{I^{\mp}_{\epsilon}=\frac{\pi}{2j+1}\brac{u\mp\bar{u}}^{2j}(-\sigma)^{\epsilon}.}
Therefore, in the regular case, the constraint for the redefined one point amplitude is
\eq{\label{RefSymmCnstr-reg}f_{\sigma}(j)=+f_{-\sigma}(-j-1).}

\section{Some Useful Formulae}
\subsection{\label{Gamma}$\Gamma$ Function Identities}
\eq{\int_0^1\mathrm{d}t\hskip .1cm t^{a-1}(1-t)^{b-1}=\frac{\Gamma(a)\Gamma(b)}{\Gamma(a+b)}}
\eq{\int_0^{\infty}\mathrm{d}t\hskip .1cm (1+t^2)^{\alpha}=\frac{\sqrt{\pi}}{2}\frac{\Gamma(-\alpha-\frac{1}{2})}{\Gamma(-\alpha)}}
\eq{\Gamma(2j)=\frac{1}{\sqrt{\pi}}(2)^{2j-1}\Gamma(j)\Gamma(j+\frac{1}{2})}
\eq{\Gamma(z)\Gamma(1-z)=\frac{\pi}{\sin(\pi z)}}

\subsection{\label{Hypergeo}Analytic Continuations of the Hypergeometric Function}
The formulae stated here are taken from \cite{Bateman}.
\spliteq{F\left(a,b;c\left\arrowvert\frac{1}{u}\right)\right.&=\frac{\Gamma(c)\Gamma(b-a)}{\Gamma(b)\Gamma(c-a)}\brac{-\frac{1}{u}}^{-a}F(a,1-c+a;1-b+a\arrowvert u)+\\
&+\frac{\Gamma(c)\Gamma(a-b)}{\Gamma(a)\Gamma(c-b)}\brac{-\frac{1}{u}}^{-b}F(b,1-c+b;1-a+b\arrowvert u).}
\spliteq{F\left(a,b;c\left\arrowvert z\right)\right.=\frac{\Gamma(c)\Gamma(c-a-b)}{\Gamma(c-a)\Gamma(c-b)}F(a,&b;a+b-c+1\arrowvert 1-z)+\\
+\frac{\Gamma(c)\Gamma(a+b-c)}{\Gamma(a)\Gamma(b)}(1-z)^{c-a-b}&F(c-a,c-b;c-a-b+1\arrowvert 1-z).}

\subsection{\label{Appell}Analytic Continuations of the Appell Function $F_1$}
The following formulae are taken from \cite{Exton}.
\spliteq{F_1(\alpha,\beta,\beta';\gamma\arrowvert u;z)&=\frac{\Gamma(\gamma)\Gamma(\gamma-\alpha-\beta')}{\Gamma(\gamma-\alpha)\Gamma(\gamma-\beta')}(1-u)^{-\beta}z^{-\beta'}\cdot\\
\cdot G_2&\left(\beta,\beta';1+\beta'-\gamma,\gamma-\alpha-\beta'\left\arrowvert\frac{u}{1-u};\frac{1-z}{z}\right)\right.+\\
&+\frac{\Gamma(\gamma)\Gamma(\alpha+\beta'-\gamma)}{\Gamma(\alpha)\Gamma(\beta')}(1-u)^{-\beta}(1-z)^{\gamma-\alpha-\beta'}\cdot\\
\cdot F_1&\left(\gamma-\alpha,\beta,\beta';\gamma-\beta-\beta'\left\arrowvert\frac{1-z}{1-u};1-z\right)\right.,}
in a neighbourhood of $(u,z)=(0,1)$. The function $G_2$ is one of Horn's functions: 
\eq{G_2(\beta,\beta';\alpha,\alpha'\arrowvert u;z):=\sum_{m,n}(\beta)_m(\beta')_n(\alpha)_{n-m}(\alpha')_{m-n}\frac{u^m}{m!}\frac{z^n}{n!},}
$(\alpha)_m:=\frac{\Gamma(\alpha+m)}{\Gamma(\alpha)}$ being the Pochhammer symbols.
\spliteq{F_1(\alpha,\beta,\beta';\gamma\arrowvert u;z)&=\frac{\Gamma(\gamma)\Gamma(\beta'-\alpha)}{\Gamma(\beta')\Gamma(\gamma-\alpha)}(-z)^{-\alpha}\cdot\\
\cdot F_1&\left(\alpha,\beta,1-\gamma+\alpha;1-\beta'+\alpha\left\arrowvert\frac{u}{z};\frac{1}{z}\right)\right.+\\
&+\frac{\Gamma(\gamma)\Gamma(\alpha-\beta')}{\Gamma(\alpha)\Gamma(\gamma-\beta')}(-z)^{-\beta'}\cdot\\
\cdot G_2&\left(\beta,\beta';1-\gamma+\beta',\alpha-\beta'\left\arrowvert -u;\frac{1}{z}\right)\right.,}
in a neighbourhood of $(u,z)=(0,\infty)$.
\spliteq{F_1(\alpha,\beta,&\beta';\gamma\arrowvert u;z)=\frac{\Gamma(\gamma)\Gamma(\gamma-\alpha-\beta-\beta')}{\Gamma(\gamma-\beta-\beta')\Gamma(\gamma-\alpha)}z^{-\alpha}\cdot\\
\cdot F_1&\left(\alpha,\beta,1-\gamma+\alpha;1+\alpha+\beta+\beta'-\gamma\left\arrowvert\frac{z-u}{z};\frac{z-1}{z}\right)\right.+\\
&+\frac{\Gamma(\gamma)\Gamma(\alpha+\beta+\beta'-\gamma)}{\Gamma(\alpha)\Gamma(\beta+\beta')}(1-z)^{\gamma-\alpha-\beta-\beta'}z^{\beta+\beta'-\gamma}\cdot\\
\cdot G_2&\left(\beta,\gamma-\beta-\beta';1+\beta+\beta',\alpha+\beta+\beta'-\gamma\left\arrowvert\frac{z-u}{1-z};\frac{1-z}{z}\right)\right.,}
in a neighbourhood of $(u,z)=(1,1)$, $\abs{u-z}\ll 1$.

\bibliographystyle{utphys}
\bibliography{draftbib}

\end{document}